\documentclass[]{aa}
\usepackage{graphicx}
\usepackage{color}
\definecolor{blue}{rgb}{0.0,0.0,1.0} 

\def\boo{$\lambda$~Bootis~}
\begin{document}
\title{On the \boo spectroscopic binary hypothesis}
\author{Ch.~St{\"u}tz, E.~Paunzen}
\mail{Ernst.Paunzen@univie.ac.at}
\institute{
  Institut f{\"u}r Astronomie der Universit{\"a}t Wien,
  T{\"u}rkenschanzstr. 17, A-1180 Wien, Austria
}
\date{Received 07 August 2006 / Accepted 29 August 2006}
\abstract{
  It is still a matter of debate if the group of \boo stars is
  homogeneously defined. A widely discussed working hypothesis formulates that two
  apparent solar abundant stars of an undetected spectroscopic binary system mimic 
  a single metal-weak spectrum preventing any reliable analysis of the group characteristics.
}{
  Is the proposed spectroscopic binary model able to explain the observed abundance
  pattern and photometric metallicity indices for the group members? What is the
  percentage of undetected spectroscopic binary systems?
}{
  We have used the newest available stellar atmospheres to synthesize 105 
  hypothetical binary systems in the relevant astrophysical parameter range. 
  These models were used to derive photometric indices. As a test, values for 
  single stellar atmospheres, Vega and two typical \boo stars, HD\,107233 and 
  HD\,204041, were generated.
}{
  The synthesized indices fit the standard lines and the observations of the
  three stars excellently. For about 90\% of the group members, the spectroscopic
  binary hypothesis can not explain the observations.
}{
  A carefully preselection of \boo stars results in a homogeneous group
  of objects which can be used to investigate the group characteristics.
}
\maketitle
%
% ---------------------------------------------------------------------------

\section{Introduction}

More than 60 years after the first notification of the peculiar nature of 
\boo (HR\,5351) itself was addressed in the spectral classification survey by 
Morgan et al. (\cite{morg43}) the origin and even the existence of a homogeneous 
group of $\lambda$\,Bootis stars is still a matter of debate.

Several theories were developed to explain the main characteristic of this
group which comprise of main sequence late B to early F type stars: the lighter 
elements (C, N, O and S) are solar abundant whereas the heavier elements are 
significantly underabundant (Paunzen \cite{pa04}).

In this letter we quantify the hypothesis formulated by Faraggiana \& Bonifacio 
(\cite{farag99}) and Gerbaldi et al. (\cite{ger03}) that some, if not all, 
$\lambda$\,Bootis stars are in fact undetected spectroscopic binary systems with
two solar abundant components simulating a combined single lined metal weak spectrum.
%
% ---------------------------------------------------------------------------

\section{The group of $\lambda$\,Bootis stars and the spectroscopic binary hypothesis} \label{lbs}

The group of \boo stars is quite outstanding
on the upper main sequence. It comprises only 2\% or less
of all objects in the relevant spectral domain and the only 
difference to normal type stars is the abundance
pattern. These stars have moderate to extreme (up to a factor 100) surface 
underabundances of most Fe-peak elements (with the exception of Na)
and solar abundances of lighter elements (C, N, O, and S). 

Our working group has tried to establish unambiguous 
membership criteria and to sort out misclassified objects 
cumulating in the list published by Paunzen et al. (\cite{pa02}). 
At this time, we believe it includes the most probable members, 
on the basis of various membership criteria, 57 in total. However,
the known spectroscopic binary systems were already excluded (see Section
2 therein). We have to emphasize that this list is
not an ``ultimate list'' but was compiled on the basis of observational
evidence, e.g. results from spectral classification 
(Paunzen \cite{pa01}), starting from the first catalogues by Renson, Faraggiana \& 
B\"ohm (\cite{re90}) and Paunzen et al. (\cite{pa97}).

The origin of the peculiar elemental abundances for the \boo group 
can be explained by selective accretion of circumstellar or interstellar material 
(Venn \& Lambert \cite{ven90}, Waters, Trams \& Waelkens \cite{wa92}, 
Kamp \& Paunzen \cite{ka02}, Andrievsky \cite{and06}).
This scenario is widely accepted to date. 
However, the issue still remains whether the group itself is homogeneously defined 
as addressed by Gerbaldi et al. (\cite{ger03}). Their working group has formulated the
hypothesis that the tendency to detect lower abundances for numerous elements in high resolution
spectroscopy abundance analysis might be explained by assuming an 
unidentified (or unidentifiable) binary system of two ``normal'' (solar abundant) 
stars with similar spectral type. As a consequence they conclude that the group of
\boo stars consists of single type objects and an unknown, but probable high 
percentage of undetected spectroscopic binary systems. 

In the following we investigate this hypothesis by comparing synthetic 
photometric indices of binary systems to those of apparent group members.
%
% ---------------------------------------------------------------------------
\begin{figure*}
\begin{center}
\includegraphics[width=160mm]{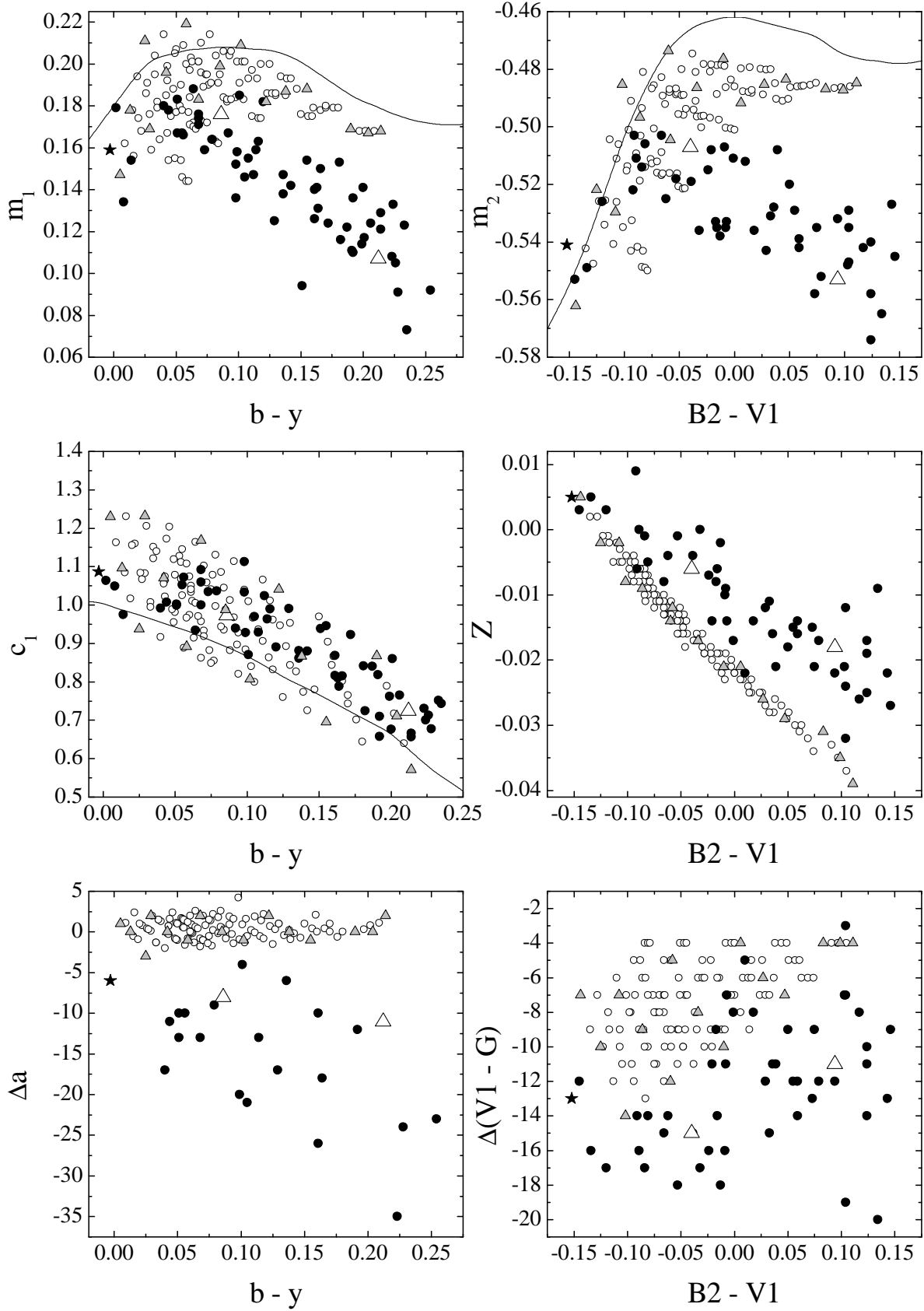}
\caption{A comparison of the synthetic Str\"omgren-Crawford $uvby\beta$, Maitzen $\Delta$a 
(left panels) and the Geneva 7 color system (right panels) with the observations for the
group of \boo stars (full circles). The open 
circles are the binary models whereas the grey shaded triangles are the 
single star models used to construct them.
Vega is marked as asterisk, the two synthetic ``standard'' \boo stars are 
open triangles. The standard lines are from the literature.}
\label{fig1}
\end{center}
\end{figure*}
% ---------------------------------------------------------------------------

\section{Modelling the spectroscopic binary systems} \label{models}

The working hypothesis is rather simple: a spectroscopic binary system
with two solar abundant components has to simulate the total energy flux 
distributions and thus the photometric colors of an apparent single 
$\lambda$\,Bootis star. 

To investigate the chances of $\lambda$\,Bootis stars being disguised
binary systems, we modelled a set of stars spanning the range in fundamental
parameters (Table\,\ref{grid}) populated by $\lambda$\,Bootis stars (Paunzen
et al. \cite{pa02}).
% ---------------------------------------------------------------------------
\begin{table}[t]
 \begin{center}
  \caption{Parameter space used for the models and the astrophysical parameters 
           of the standard star Vega and the \boo stars HD\,107233 and HD\,204041.}
  \begin{tabular}[h]{lcccc}
    \hline
    \hline
                      &      Models     & Vega & 107233  & 204041 \\
    \hline
    $T_{\rm eff}$ [K] & 7000  $-$ 9000  & 9560 &  6900   &  8100   \\
    $\log g$ [cgs]    & 3.50  $-$ 4.50  & 4.05 &  3.80   &  4.10   \\
    $v_{turb}$ [km\,s$^{-1}$] & 2       & 2    &  3      &  2      \\
    Convection        & \multicolumn{4}{c}{Canuto \& Mazzitelli (\cite{cm91})} \\
    \hline
    \hline
  \end{tabular}
  \label{grid}
  \end{center}
\end{table}
% ---------------------------------------------------------------------------
The turbulent velocity was fixed at a value of 2\,km\,s$^{-1}$, convection
was modelled according to Canuto \& Mazzitelli (\cite{cm91}, hereafter CM) 
since we are situated in the region of weak to no convection. 
To model the atmospheres as well as the energy flux distribution of these 
stars and of our photometric anchor Vega, we used the code LLmodels\,v.SE/8.0 
(Shulyak et al. \cite{llm04}). This model atmosphere code supports individual
chemical composition, VCS theory (Vidal, Cooper \& Smith \cite{vcs73}, 
Lemke \cite{lemke97}) for the treatment of Hydrogen lines and the CM
convection treatment.
To represent the atomic line opacities we created a separate linelist for
 each model. The line parameters where obtained form the Vienna Atomic Line Database
 (VALD, Kupka et al. \cite{vald}). All lines for which 
 $ l_{\nu} / \alpha_{\nu} \ge 1\% $ 
 ($l_{\nu}$ and $\alpha_{\nu}$ are line and continuum absorption coefficient)
 was realized for the given model structure were selected.
 The step size in wavelength of our synthetic spectra was set to 0.1\,{\AA}.

We built energy flux distributions of 105 hypothetical binary stars by preparing
all possible combinations of always two spectra. On these, we performed synthetic 
photometry for the 
Str\"omgren-Crawford $uvby\beta$, Maitzen $\Delta$a (Paunzen, St\"utz \& Maitzen 
\cite{pa05}) and the Geneva 7-color system.
As it is common practice, Vega was our reference point for the synthetic photometry.
We modelled this stars atmosphere according to the parameters of 
Hill \& Landstreet (\cite{hill93}), the observed photometric indices were 
inquired from the {\sc{SIMBAD} and \sc{GCPD}} astronomical databases.

To show also a comparison to synthesized intrinsic $\lambda$\,Bootis stars
we modelled, the stars HD\,107233 and 
HD\,204041 with the parameters and abundances
listed in Heiter, Weiss \& Paunzen (\cite{he02}).

For our statistical analysis we have not taken the reddening into account.
Gerbaldi et al. (\cite{ger03}) have extensively discussed the problems 
using the standard Str\"omgren-Crawford $uvby\beta$ dereddening procedures for the
group of \boo stars. We have compared their values with those of
Paunzen et al. (\cite{pa02}) and found a good agreement. However, the reddening for
most objects is negligible because they are all located in the solar neighbourhood. 

% ---------------------------------------------------------------------------
\begin{table}[t]
 \begin{center}
  \caption{Location of the stars according to Figure\,\ref{fig1}. A plus sign means
  that the objects are located within the area of the spectroscopic binary models.
  The results for the individual stars is listed in Table \ref{result}.}
  \begin{tabular}[h]{lcccccc}
   \hline
   \hline
   Diagram & (\,1\,) & (\,2\,) & (\,3\,) & (\,4\,) & (\,5\,) & (\,6\,) \\
   $\Delta$a             & $-$ & $-$ & $-$   &       &   &       \\
   $\Delta$(V1\,-\,G)/Z  & $-$ & $-$ & +/$-$ & +/$-$ & + &       \\
   m$_1$/m$_2$           & $-$ &     & $-$   & $-$   & + &       \\
   m$_1$                 &     &     &       &       &   & +/$-$ \\
   \hline
   \hline
  \end{tabular}
  \label{legend}
  \end{center}
\end{table}
% ---------------------------------------------------------------------------
%

\section{Results and Conclusions}

Figure\,\ref{fig1} shows the results of the relevant photometric diagrams. 
The observational data for the \boo stars were taken from Paunzen et al. \cite{pa02}. 
The classical metallicity sensitive diagrams m$_1$ versus (b\,-\,y) and 
m$_2$ versus (B2\,-\,V1) show that for cooler objects, the \boo stars are nicely 
separated from the models on the main sequence (see c$_1$ versus (b\,-\,y)). 
These observational facts were also described in Paunzen et al. (\cite{pa97}).

% ---------------------------------------------------------------------------
\begin{table}[h]
 \begin{center}
  \caption{The results for $\lambda$ Bootis stars taken from Paunzen et al. \cite{pa02}
  according to Figure\,\ref{fig1} and Table \ref{legend}. We would only consider the stars of group (\,5\,) 
  as being good candidates for undetected spectroscopic binary systems. 
  No conclusion can be drawn for group (\,6\,), the abbreviations are {\bf{P}}assed or {\bf{F}}ailed. 
  The last row contains the number of objects as part of the total sample.}
  \begin{tabular}[h]{rrrrrr}
   \hline
   \hline
   \multicolumn{1}{c}{(\,1\,)} & \multicolumn{1}{c}{(\,2\,)} & \multicolumn{1}{c}{(\,3\,)} & 
   \multicolumn{1}{c}{(\,4\,)} & \multicolumn{1}{c}{(\,5\,)} & \multicolumn{1}{c}{(\,6\,)} \\
      6870 &     319 &   31295 &  111604 &   23392 &   13755\,{\bf{P}} \\
     11413 &    7908 &   35242 &  156954 &  110377 &   15165\,{\bf{P}} \\
     24472 &   30422 &  101108 &  193256 &  170680 &   54272\,{\bf{P}} \\
     75654 &   74873 &         &  193281 &  198160 &   87271\,{\bf{P}} \\
     81290 &   91130 &         &         &         &   90821\,{\bf{F}} \\
     83041 &  110411 &         &         &         &  105759\,{\bf{P}} \\
     83277 &  125162 &         &         &         &  111005\,{\bf{P}} \\
     84123 &  130767 &         &         &         &  120500\,{\bf{F}} \\
    102541 &  183324 &         &         &         &  120896\,{\bf{P}} \\
    105058 &  204041 &         &         &         &  175445\,{\bf{F}} \\
    106223 &  221756 &         &         &         &         \\
    107233 &         &         &         &         &         \\
    109738 &         &         &         &         &         \\
    125889 &         &         &         &         &         \\
    142703 &         &         &         &         &         \\
    142994 &         &         &         &         &         \\
    149130 &         &         &         &         &         \\
    153747 &         &         &         &         &         \\
    154153 &         &         &         &         &         \\
    168740 &         &         &         &         &         \\
    168947 &         &         &         &         &         \\
    184779 &         &         &         &         &         \\
    192640 &         &         &         &         &         \\
    210111 &         &         &         &         &         \\
    216847 &         &         &         &         &         \\
   \hline
    44\,\% & 19\,\% & 5\,\% & 7\,\% & 7\,\% & 18\,\% \\
   \hline
   \hline
  \end{tabular}
  \label{result}
  \end{center}
\end{table}
% ---------------------------------------------------------------------------
%

Very interesting are the other three diagrams which include directly metallicity 
depend indices (Paunzen et al. \cite{pa05}). The most important index is $\Delta$a, for
which, unfortunately, the least measurements are available whereas $\Delta$(V1\,-\,G) 
seems less sensitive with a larger scatter. We notice that Vega as
metal-deficient object (Hill \& Landstreet \cite{hill93}, Garc\'ia-Gil et al. \cite{gar05}) 
shows, as expected, a reduced $\Delta$a value. Otherwise, the standard lines and observations
of Vega fit the synthetic values very good. The differences of the synthetic and observed
values for the two \boo stars are only 1 to 2\,mmag for $\Delta$a, Z, and 
$\Delta$(V1\,-\,G).

According to Figure\,\ref{fig1} we have divided our sample in six different groups numbered
from (\,1\,) to (\,6\,) depending if the placement is compatible with the spectroscopic 
binary models taking an error of $\pm$5\,mmag into account. 
The final division is listed in Table \ref{legend} whereas
Table\,\ref{result} shows the results for the individual stars. 
The first two groups include objects which are not located in the spectroscopic
binary area (SBA hereafter) in any diagram. 
The third and fourth group include stars which are situated in the SBA either in
the $\Delta$(V1\,-\,G) or Z diagram, but not in the classical metallicity and the $\Delta$a diagram 
or do not have an available $\Delta$a measurement (fourth group).
The fifth group comprise good candidates for undetected spectroscopic binaries among the \boo 
group because these stars compare well to the synthetic photometry of the binary models. 
HD\,198160 is already known as close visual binary star whereas inconsistent $v \sin i$ 
measurements were reported by Heiter et al. (\cite{he02}) for HD\,170680. The four objects 
of this group certainly deserve further attention in the future.
The last group includes those stars which could only be tested within 
the m$_1$ versus (b\,-\,y) diagram which prevents any clear conclusion.

If we compare the list of Table\,\ref{result} with the results of Faraggiana et al. 
(\cite{farag04}), only two stars, HD\,11413 and HD\,210111, were reported as spectroscopic 
binary systems on the basis of radial velocity shifts of two and three spectra, respectively. 
Both are well investigated $\delta$ Scuti pulsators (Koen et al. \cite{ko03}, 
Breger et al. \cite{br06}) which makes them especially interesting for further investigations. 
We conclude that if these stars are indeed binary systems, at least one component is of 
\boo type similar to the systems investigated by Iliev et al. (\cite{il02}).

The percentage of undetected spectroscopic binary systems mimicking a single, metal-weak object
seems very low. From 47 well investigated stars, groups (\,1\,) to (\,5\,), only four objects 
seem good candidates for a further investigation which is below 10\% of the complete sample.

A carefully preselection of \boo stars results in a homogeneous group of intrinsic \boo stars 
which can be used to investigate the group properties in more detail.
%
% ---------------------------------------------------------------------------
\begin{acknowledgements}
This research was performed within the projects  {\sl P17580}, {\sl P17890} and {\sl P17920} 
of the Austrian Fonds zur F{\"o}rderung der wissen\-schaft\-lichen Forschung (FwF). 
Use was made of the SIMBAD database, operated at the CDS, Strasbourg, France, the
NASA's Astrophysics Data System and the General Catalogue of Photometric Data (GCPD). 
\end{acknowledgements}
% ---------------------------------------------------------------------------

% ---------------------------------------------------------------------------
\end{document}